\documentclass[10pt,amsmath,amssymb,aps,twocolumn]{revtex4}
\usepackage{amsmath}
\usepackage{amssymb}
\usepackage{graphics}
\usepackage{graphicx}
\newcommand {\be}{\begin{equation}}
\newcommand {\ee}{\end{equation}}
\newcommand {\mb}[1]{\mathbf{#1}}
\begin{document}
\unitlength=1mm
\title{Green's Function Approach to the Bose-Hubbard Model}

\date{\today}
\author{Matthias Ohliger${}^1$ and Axel Pelster${}^{2,3}$}
\affiliation{${}^1$Institut f\"ur Theoretische Physik, Freien Universit\"at Berlin, Arnimallee 14, 14195 Berlin, Germany\\
${}^2$Fachbereich Physik und Forschungszentrum OPTIMAS, Technische Universit\"at Kaiserslautern,
Erwin-Schr\"odinger Stra{\ss}e, Geb\"aude 46, 67663 Kaiserslautern, Germany\\
${}^3$Hanse-Wissenschaftskolleg, Lehmkuhlenbusch 4, 27753 Delmenhorst, Germany}

\begin{abstract}
We use a diagrammatic hopping expansion to calculate finite-temperature Green functions of the Bose-Hubbard model 
which describes bosons in an optical lattice.
This technique allows for a summation of subsets of diagrams, so the divergence of the Green function leads to non-perturbative 
results for the boundary between the superfluid and the Mott phase for finite temperatures. Whereas the first-order calculation 
reproduces the seminal mean-field result, the second order goes beyond and shifts the phase boundary in the immediate vicinity 
of the critical parameters determined by high-precision Monte-Carlo simulations of the Bose-Hubbard model. In addition, our 
Green's function approach allows for calculating the excitation spectrum both for zero and finite temperature and for 
determining the effective masses of particles and holes. 
\end{abstract}
\maketitle

\section{Introduction} 

Ultracold bosonic gases trapped in the periodic potential of optical lattices represent tunable
model
systems for studying the physics of quantum phase transitions \cite{fisher,jaksch,bloch2,sachdev}. They are described by
the Bose-Hubbard Hamiltonian which decomposes into two parts: $\hat{H}=\hat{H}_0+\hat{H}_1$. The first term
$\hat{H}_0=\sum_i\left[ U \hat{n}_i(\hat{n}_i-1)/2-\mu\hat{n}_i \right]$,
with the on-site energy $U$ and the chemical potential $\mu$, describes the repulsion of more than one boson
residing on a lattice site. It is local and diagonalizable in the occupation number basis for any lattice site.
The second term $\hat{H}_1=-J_{i,j}\sum_{i,j}\hat{a}_i^\dag\hat{a}_j$ with the hopping matrix element
$J_{i,j}=J$ if the lattice sites $i$ and $j$ are nearest neighbors and $J=0$ otherwise
describes the hopping between two sites due to the quantum-mechanical tunneling effect. The competition between the two
energy scales $U$ and $J$ determines the existence of two different phases. When the on-site energy is small compared to the
hopping amplitude, the ground state is superfluid (SF) as the bosons are delocalized in a phase coherent way over the whole
lattice. In the opposite case, where the on-site interaction dominates over the hopping term, the ground state
is a Mott insulator (MI) where each boson is trapped in one of the respective potential minima.\\
This characteristic quantum phase transition of the Bose-Hubbard model has been
studied extensively both with analytical \cite{monien,monienpade,stoof1,stoof2,santos,Holthaus5,Holthaus6} and numerical 
\cite{monienDMRG,QMC3D,QMC2D2}  methods for zero temperature, while less literature exists on the finite-temperature 
properties of this transition \cite{schmidtPRL,buonsante,pelster}. In this letter we work out an analytical Green's 
function approach in order to determine the MI-SF phase boundary and the excitation spectrum in the Mott phase
both for zero and finite temperature with a hopping expansion. Our findings compare well with the latest findings
of Quantum Monte Carlo simulations and allow to propose a thermometer for bosons in optical lattices.
In the following we restrict ourselves to a spatially homogeneous system and neglect the effects arising from the 
additional harmonic confining potential, which is present in all experimental settings, like the formation of a shell 
structure \cite{bloch1}. However, these effects could be taken into account by applying the local density approximation 
where the external potential is taken into account in form of a spatially dependent chemical potential.\\
\section{Green's Function Approach}

As all single-particle properties of a quantum many-body system are contained in its 
Green function, we base our calculation on this quantity. Because we are interested in describing a system at non-zero 
temperature, we use the imaginary-time formalism \cite{abrikosov,zinnjustin}. 
Therein, the single-particle Green function is defined as the thermal 
average of the time-ordered product of a creation and an annihilation operator in Heisenberg representation
\begin{equation}
G_1(\tau',j'|\tau,j)= {\rm{Tr}}\left\{ \frac{e^{-{\beta} \hat{H}}}{\cal Z} \hat{T}
\left[\hat{a}_{j,\rm H}(\tau)\hat{a}_{j',\rm H}^{\dag}(\tau')\right]\right\} \, ,
\end{equation}
with $\hbar=1$, $\beta=1/k_BT$ and we have introduced the partition function ${\cal Z}={\rm Tr}\{e^{-\beta\hat{H}}\}$. 
Because it is not possible to obtain analytic expressions for the eigenstates and eigenenergies of the full Bose-Hubbard 
Hamiltonian, we cannot calculate the Green function exactly. Instead, we aim at a perturbative treatment and calculate 
this quantity as a power series in the hopping matrix element $J_{i,j}$. As that parameter is small for the
Mott phase, where the lattices are deep and the interaction between particles is strong, 
we refer to this treatment as a strong-coupling expansion. In order to employ this 
perturbative expansion for finite temperature we make use of the Dirac interaction 
picture and write the imaginary-time evolution operator in form of a Dyson series,
\begin{equation}
\label{eq:UD}
\hat{U}(\tau,\tau_0)=\hat{T}\exp\left[-\int_{\tau_0}^\tau d\tau_1\hat{H}_{1}(\tau_1)\right]\,,
\end{equation}
where the time dependence of the Dirac-picture operators is determined by the local Hamiltonian $\hat{H}_0$. 
With its help we can write the Green function as  
\begin{equation}
\label{eq:g1fst}
\hspace*{-2mm}G_1(\tau ',i'|\tau,i)= {\rm Tr}\left\{ \frac{e^{-\beta\hat{H}_0}}{\cal Z}\hat{T}\left[\hat{a}^\dag_{i'}(\tau ')
\hat{a}_i(\tau)\hat{U}(\beta,0)\right]\right\},
\end{equation}
where the time-ordering operator acts also on the time variables which are resulting from the expansion of the 
Dirac imaginary-time evolution operator in (\ref{eq:UD}). 
When we now expand perturbatively in powers of the tunneling matrix element $J$, the $n-1$th order contribution
$G_1^{(n-1)}$ in (\ref{eq:g1fst}) turns out to depend on
the $n$-particle Green function of the \textit{unperturbed} system as 
$G_n^{(0)}(\tau_1',i_1';\dots;\tau_n',i_n'|\tau_1,i_1;\dots;\tau_n,i_n)=
\left\langle\hat{T}\left[\hat{a}_{i'_1}^\dag(\tau_1')\hat{a}_{i_1}(\tau_1)\dots\hat{a}_{i'_n}^\dag
(\tau_n')\hat{a}_{i_n}(\tau_n)\right]\right\rangle^{(0)}$.
In order to determine
$G_{n}^{(0)}$ we cannot use standard Wick's theorem because the unperturbed Hamiltonian $\hat{H}_0$ is not quadratic in 
the Bose operators. 
Although the lack of Wick's theorem in the present situation prevents us from using 
the powerful perturbative technique based on standard Feynman diagrams, 
we can nevertheless simplify $G_{n}$ by decomposing it 
into cumulants. To this end, we follow an approach reviewed by Metzner \cite{metzner} in the context of the Hubbard 
model which describes electrons in a conductor. This decomposition is based on the important observation that the on-site
Hamiltonian $\hat{H}_0$ is local. Consequently, the unperturbed Green functions $G_n^{(0)}$ are also local and can
be decomposed into time-dependent cumulants $C_n^{(0)}$. For instance, we have
$G_1^{(0)}(\tau ',i'|\tau,i)=\delta_{i,i'}C_1^{(0)}(\tau '|\tau)$ with the cumulant
\begin{eqnarray}
&& \hspace*{-0.4cm} C_1^{(0)}(\tau '|\tau)=\frac{1}{{\cal Z}^{(0)}}\sum_{n=0}^\infty 
 \left[\Theta(\tau-\tau ')(n+1)\,e^{(E_n-E_{n+1})(\tau-\tau ')} \right. \nonumber \\&&
\left. +\Theta(\tau'-\tau)\,n\,e^{(E_n-E_{n-1})(\tau'-\tau)}\right]\,e^{-\beta E_n} \, ,
\label{eq:C1}
\end{eqnarray}
where ${\cal Z}^{(0)}=\sum_n e^{- \beta E_n}$ is the unperturbed partition function for a single-site system
with the on-site energy eigenvalues $E_n=U n (n-1)/2 - \mu n$.
With this decomposition, we can represent $G_1^{(n)}$ diagrammatically: 
We denote an $n$-particle cumulant at a 
lattice site by a vertex with $n$ entering and $n$ leaving lines with imaginary-time variables, so we have, for instance,
for the first two cumulants
\begin{eqnarray}
\raisebox{-4mm}{\includegraphics[width=15mm]{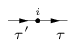}}&=&\,C_1^{(0)}(\tau'|\tau)\,, \\
\raisebox{-5mm}{\includegraphics[width=20mm]{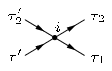}}&=&\,C_2^{(0)}(\tau_1',\tau_2'|\tau_1,\tau_2)\,.
\end{eqnarray}
Furthermore, the hopping matrix element 
is symbolized by a line connecting two vertices:
\begin{eqnarray}
{}_i\,\raisebox{0mm}{\includegraphics[width=15mm]{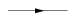}}\, {}_j&=&J_{ij} \,.
\end{eqnarray}
With all this, we can set up the diagrammatic rules for calculating the $n$th order contribution of the Green function in $J$. 
First: Draw all possible combinations of vertices with total $n$ internal and one entering and one leaving line. 
Second: Connect them in all possibles ways and assign time variables and hopping matrix elements to the lines. 
Third: Sum all site indices over all internal lattice sites and integrate all internal time variables from $0$ to $\beta$.\\
We also note here that we have to sum all site indices over the \textit{whole} lattice, no matter whether two sites 
in a diagram coincide or not. We make use of the translational invariance in 
imaginary time and transform all expressions to Matsubara space. In the \textit{second} diagrammatic rule the integrals 
over the time variables have to be replaced by sums over all bosonic Matsubara frequencies $\omega_m=2\pi m/\beta$ with 
integer $m$ where the sum of the incoming frequencies must equal the sum of the outgoing ones.\\
The formalism developed so far allows for calculating the Green function to any given order in 
the tunneling matrix element $J$. But because one 
of our main goals is to describe the phase transition between the Mott insulator and the superfluid phase, and it is 
well known in the theory of critical phenomena
that such a transition is characterized by diverging long-range correlations \cite{zinnjustin,verena}, we must 
employ a non-perturbative method which is archieved by summing an infinite subset of diagrams. In 
order to perform this task, we introduce the sum of all one-particle irreducible diagrams including their respective symmetry 
factors and multiplicities as
\begin{equation}
\label{dia:dysdia}
{\cal C}_1(\omega_m,\mathbf{k})=
\,\raisebox{0mm}{\includegraphics[width=16mm]{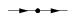}}\,+\, \raisebox{0mm}{\includegraphics[width=18mm]{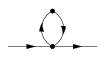}}
\,+ \ldots \,.
\end{equation}
The self-energy which describes the movement of a single particle in a many-body enviroment is defined as 
$\Sigma=1/G^{(0)}_1-1/G_1$ \cite{abrikosov}. For our case, it can be written in the 
form $\Sigma(\omega_m,\mb{k})=1/G_1^{(0)}(\omega_m)-1/{\cal C}_1(\omega_m,\mb{k})+J(\mb{k})$, 
where $J(\mb{k})=2J\sum_{\nu=1}^D\cos(k_\nu a)$ is the $D$-dimensional lattice dispersion. 
For instance, the first order in the self-energy which corresponds to a summation of all simple chain diagrams, 
yields $\tilde{G}_1(\omega_m,\mb{k})=\left[1/G_1^{(0)}(\omega_m)-J(\mb{k})\right]^{-1}$ with 
$C_1^{(0)}(\omega_m)=\sum_n\bigl[
(n+1)/(E_{n+1}-E_n+i\omega_m)-$ $n/(E_n-E_{n-1}+i\omega_m)\bigr]e^{-\beta E_n} /{\cal Z}^{(0)}$.   
We note that performing the limit 
$\tau'\searrow\tau$ in the Green function allows to obtain the quasi-momentum distribution needed to explain 
experimental time-of-flight pictures \cite{blochvisibility,alexpaper,bloch3}.\\
\section{Phase Boundary} 

The boundary of the phase transition is 
characterized by diverging long-range correlations \cite{zinnjustin,verena}. Thus, we set $\mb{k}=0$ and solve for the value of $J$ 
where the Green function diverges which is only possible for 
$\omega_m=0$. This yields up to first order in $J$
\be
\label{eq:phaseMF}
2DJ_c=\frac{\sum_ne^{-\beta E_n}}{\sum_n e^{-\beta E_n}\left(\frac{n+1}{E_{n+1}-E_n}-\frac{n}{E_n-E_{n-1}}\right)}
\ee
which coincides with the finite-temperature mean-field result \cite{pelster,buonsante,gerbier}. The zero-temperature 
limit of (\ref{eq:phaseMF}), i.e. $\label{eq:phase0}
2DJ_c=1/\left(\frac{n+1}{E_{n+1}-E_n}-\frac{n}{E_n-E_{n-1}}\right)$, agrees with the seminal mean-field result of 
Ref.~\cite{fisher}. The reason for this agreement is that each approximation becomes exact in the limit of infinite 
spatial dimension. In order to see that one must suitably scale the hopping parameter \cite{vollhardtferm,dmft1,vollhardtneu}. 
When we define $\tilde{J}=2DJ$, the contribution of the $k$th order chain diagram is proportional to 
$\tilde{J}^k$ because there exist $2D$ possibilities in a chain diagram
for every hopping line to connect to neighboring sites. The lowest-order term neglected by that summation is the one-loop 
diagram in (\ref{dia:dysdia})
which has two internal lines but only one free index and is, therefore, proportional to $2DJ^2=\tilde{J}^2/(2D)$. Thus, it 
vanishes in the limit of $D\rightarrow\infty$.\\
Taking now the one-loop diagram into account yields
\be
\label{eq:G2tilde}
\tilde{G}^{(2)}_1(\omega_m,\mb{k})=\frac{1}{1/G_1^{(0)}(\omega_m)-J(\mb{k})+\Sigma^{(2)}(\omega_m,\mb{k})}\,,
\ee
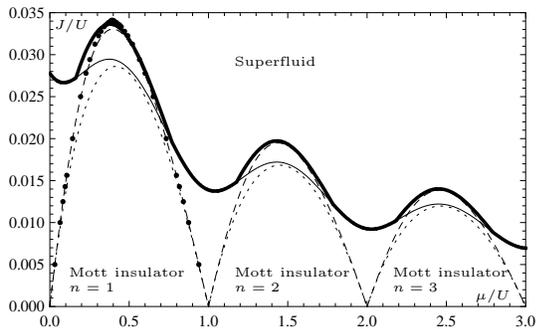
\begin{figure}[t]
\unitlength1mm
\begin{picture}(70,45)
\put(62,4){\tiny $\mu/U$}
\put(6,40){\tiny $J/U$}
\put(30,35){\tiny Superfluid}
\put(8,5){\tiny $n=1$}
\put(8,7){\tiny Mott insulator}
\put(30,5){\tiny $n=2$}
\put(30,7){\tiny Mott insulator}
\put(51,5){\tiny $n=3$}
\put(51,7){\tiny Mott insulator}
\includegraphics[width=7cm]{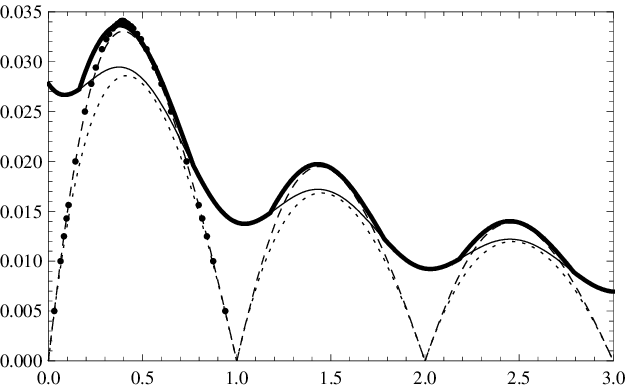}
\end{picture}
\caption{\label{phase}Quantum phase diagram for $D=3$. Thick (solid): Second (first) order, $k_BT=0.1\,U$. 
Dashed (dotted): Second (first) order, $T=0$. Dots: QMC Data for $T=0$ \cite{QMC3D}.}
\end{figure}
\hspace*{-3mm}where $\Sigma^{(2)}(\omega_m,\mb{k})=G_1^{(2B)}(\omega_m)/\left[G_1^{(0)}(\omega_m)\right]^2$ is the second-order self-energy with the value of the one-loop diagram $G_1^{(2B)}(\omega_m)$. The analytic formula for the phase 
boundary resulting from this formula is shown in Fig. \ref{phase}. This result coincides in the limit of $T\to 0$ 
with the effective potential formalism employed in Ref.~\cite{santos}. We want to emphasize that, unlike the first order 
(\ref{eq:phaseMF}), the one-loop corrected result depends on the system dimension in an non-trivial way and that the 
corrections are larger in two than in three dimension which is consistent with the fact that the first-order results 
becomes exact in the limit $D\to\infty$. Note that even higher-order corrections for the effective potential formalism have been obtained in Refs.~\cite{Holthaus5,Holthaus6},
which turned out to be indistinguishable from the Quantum Monte-Carlo data in Ref.~\cite{QMC3D}.

For finite temperature, the phase boundary is shifted towards larger values of $J_c$ 
as thermal fluctuations suppress quantum correlations which are responsible for the formation of the superfluid. This effect, 
which occurs both in first and in second order, is most important between the Mott lobes as fluctuations are strongest when 
the average particle number is not near an integer value. The correction to the mean-field result arising from the one-loop 
diagram, visible in the difference between first and second order curve, plays an important role only near the tip of the 
Mott lobe. This feature stems from the fact that \textit{quantum} fluctuation are notably increased when the system 
approaches the quantum critical point at the tip of the lobe \cite{sachdev}. Thus, we can say the quantum phase diagram 
consists of a thermally dominated region where the influence of thermal fluctuations is large and a quantum dominated region 
where the quantum corrections from the one-loop diagram are most important. \\
\begin{figure}[t]
\unitlength1mm
\begin{picture}(70,45)
\put(65,4){\tiny $k$}
\put(6,41){\tiny $\omega_{\rm pair}/U$}
\includegraphics[width=7cm]{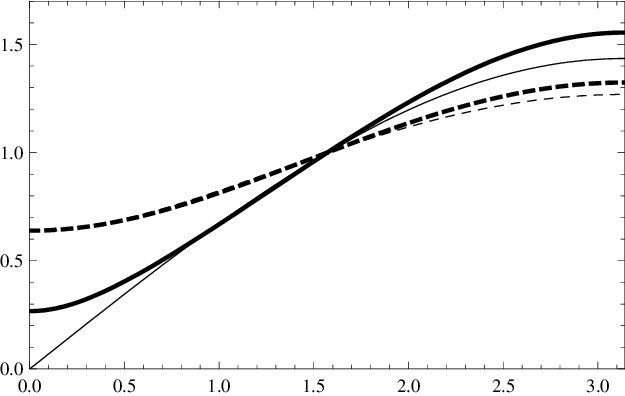}
\end{picture}
\caption{\label{speccomp}Dispersion relation of the pair-excitations for $T=0$. Thick lines: Second order. 
Thin lines: First order. Solid: $J=0.029\,U$. Dashed: $J=0.025\,U$.}
\end{figure}
\section{Excitation Spectrum} 

\begin{figure}[t]
\unitlength1mm
\begin{picture}(70,45)
\includegraphics[width=7cm]{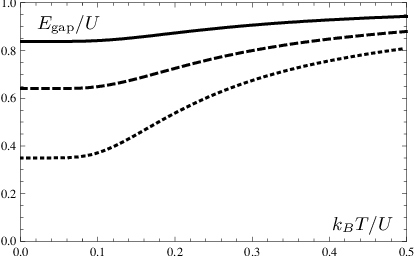}
\end{picture}
\caption{\label{gap} Temperatue dependence of gap for unit filling. Solid: $J=0.008\,U$. 
Dashed: $J=0.012\,U$. Dotted: $J=0.025\,U$.}
\end{figure}

\begin{figure}[t]
\unitlength1mm
\begin{picture}(70,44)
\put(58,4){\tiny $J/U$}
\put(9,41){\tiny  $J M_{\rm eff}/U$}
\includegraphics[width=7cm]{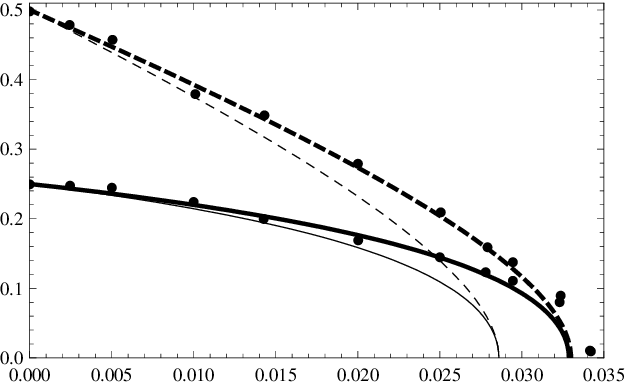}
\end{picture}
\caption{\label{masses} Effective masses of quasi-particles (solid lines) and quasi-holes (dashed lines) for $T=0$. 
Thick lines: Second order. Thin lines: First order. Dots: QMC data \cite{QMC3D}.}
\end{figure}

For a system in the Mott phase three different types of excitations exist. 1. 
The addition of a particle from the environment (particle excitation). 2. The removal of a particle (hole excitation). 3. 
The creation of a particle-hole pair (pair excitation). The last one is most important from a physical point of view 
because it is also possible in an isolated system as realized in the current experiments. In the superfluid phase, 
there are additional excitations corresponding to fluctuations of the phase of the superfluid order parameter. However, 
these features can not be investigated within the present formalism but with the related effective action method \cite{barry}.\\
In order to obtain the spectrum of the quasi-particles, which is given by the poles of the real-time Green function 
for finite temperature we must analytically continue our imaginary-time result. This is achieved by performing the 
replacement $i\omega_m\rightarrow\omega+i\eta$ with $\eta\to 0$ \cite{abrikosov}. The first and second order, respectively, 
yield excitation spectra of which the former one agrees in the limit $T\to 0$ with the mean-field result from
 Ref.~\cite{stoof1}. In the following we restrict ourselves to the most interesting case $D=3$. Both finite 
temperature and one-loop corrections are most effective for small wave numbers, the former because the effect of temperature 
is the suppression of quantum correlations which mainly exist for long wavelengths, the latter because the dominant 
fluctuations near a quantum critical point are the ones with vanishing wave number as shown in Fig. \ref{speccomp}\\
All spectra show a characteristic gap which vanishes at the critical point, i.e. at the value of $J_c$ at the tip of a 
Mott lobe. In Fig. \ref{gap} its temperature dependence is shown. As this gap is a quantity which is experimentally accessible 
 \cite{bloch1}, it could serve as a method to determine the temperature of bosons in an optical lattice. 
The quasi-particle can be ascribed an effective mass which is shown in Fig. \ref{masses} for the particle- and for the 
hole-excitations. They both become massless at the critical point which is a result of the U(1) symmetry breaking at the 
second-order phase transition. \\
\section{Conclusion and Outlook}

We have presented a powerful formalism to calculate the Green function for the 
Bose-Hubbard model in the Mott phase. It allowed us to improve the mean-field phase boundary in an analytic way both for 
finite and zero temperature where the former result deviates from recent Quantum Monte-Carlo studies by only 3\% for $D=3$ 
in Ref.~\cite{QMC3D}.  For finite temperature first analytic results beyond mean-field theory have been 
presented and the importance of both  thermal and quantum fluctuations in the different regions of the phase diagram has been 
clarified. In addition, we have calculated the excitation spectrum of the quasi-particles and derived its effective masses 
and compared them to numerical findings. We have also investigated the characteristic energy gap and determined its 
temperature dependence. 
More applications of this approach have recently allowed in Ref.~\cite{Grass} to obtain more insight 
into the superfluid phase by combining the present technique with the effective action approach from Ref.~\cite{barry}.\\
\section*{Acknowledgement} 

We cordially thank B. Bradlyn, H. Enoksen, A. Hoffmann, H. Kleinert, F. Nogueira, R. Graham, 
and F.E.A. dos Santos for stimulating and helpful discussions. 

\end{document}